\documentclass[twocolumn,showpacs,nofootinbib,prl,floatfix]{revtex4}
\usepackage{graphicx}
\usepackage{amsmath}
\usepackage{amssymb}
\usepackage{textcomp}
\usepackage{color}
\usepackage{hyperref}

\begin{document}

\title{Measurement of Linear Stark Interference in $^{199}$Hg}
\author{T.~H.~Loftus}
\altaffiliation{Present address: AOSense, Inc., Sunnyvale CA 94085}
\author{M.~D.~Swallows}
\altaffiliation{Present address: JILA, University of Colorado, and NIST, Boulder, CO 80309}
\author{W.~C.~Griffith}
\altaffiliation{Present address: Subatomic Physics Group, Los Alamos National Laboratory, Los Alamos, NM 87545}
\author{M.~V.~Romalis}
\altaffiliation{Permanent address: Department of Physics, Princeton University, Princeton, NJ 08544}
\author{B.~R.~Heckel}
\author{E.~N.~Fortson}
\affiliation{Department of Physics, Box 351560, University of
Washington, Seattle, WA 98195-1560}
\date{\today}

\newcommand{\Hg}{\ensuremath{^{199} \textrm{Hg}}}
\newcommand{\Hgline}{\ensuremath{6^{1}{S}_{0} \rightarrow 6^{3}{P}_{1} }}
\newcommand{\cred}{\color{red}}
\newcommand{\cblue}{\color{blue}}
\newcommand{\cblack}{\color{black}}
\newcommand{\cgreen}{\color{green}}

\begin{abstract}
We present measurements of Stark interference in the 6$^1S_0$ $\rightarrow$ 6$^3P_1$ transition in $^{199}$Hg, a process whereby a static electric field $E$ mixes magnetic dipole and electric quadrupole couplings into an electric dipole transition, leading to $E$-linear energy shifts similar to those produced by a permanent atomic electric dipole moment (EDM). The measured interference amplitude, $a_{SI}$ = $(a_{M1} + a_{E2})$ = (5.8 $\pm$ 1.5)$\times 10^{-9}$ (kV/cm)$^{-1}$, agrees with relativistic, many-body predictions and confirms that earlier central-field estimates are a factor of 10 too large. More importantly, this study validates the capability of the $^{199}$Hg EDM search apparatus to resolve non-trivial, controlled, and sub-nHz Larmor frequency shifts with EDM-like characteristics.

\end{abstract}


\maketitle

By mixing opposite-parity states, a static electric field $E$ can induce an electric dipole ($E1$) amplitude in a magnetic dipole ($M1$) transition; one of the resulting Stark interference (SI) effects enables precise measurements of atomic parity nonconservation (e.g., Refs. \cite{Bouchiat75, Wood97}). Analogous effects occur when an $E$-field mixes $M1$ and electric quadrupole ($E2$) amplitudes into an $E1$ transition. This type of SI, first identified \cite{Hodgdon91} and observed \cite{Chen94} with Rb and later calculated for Hg \cite{Lamoreaux1992,Beloy2009}, is relevant to searches for permanent atomic electric dipole moments (EDMs)\cite{Khriplovich1997}: SI is similar to a finite EDM as both produce $E$-linear ground-state energy shifts that depend on the electronic or nuclear spin.  Thus SI is a potential source of systematic error in EDM searches and also serves as a useful proxy for EDM effects.


We have measured SI in the $^{199}$Hg 6$^1S_0$ $\rightarrow$ 6$^3P_1$ transition, the 254 nm $E1$ intercombination line used to search for the EDM of $^{199}$Hg \cite{Griffith2009,Swallows2009}. To our knowledge, this is the first observation of SI in a diamagnetic atom, where the ground-state polarization is specified by the nuclear spin. The measured SI amplitude agrees with relativistic, many-body predictions \cite{Beloy2009}, and confirms that earlier central-field estimates \cite{Lamoreaux1992} were a factor of 10 too large and thus overestimated SI systematic errors in $^{199}$Hg EDM searches.  In addition, this study was conducted with the $^{199}$Hg EDM search apparatus, and demonstrates its capability to resolve non-trivial, controlled, and sub-nHz Larmor frequency shifts with EDM-like signatures.

 $^{199}$Hg has nuclear spin $I = 1/2$ and thus the 6$^{3}{P}_{1}$ state is a hyperfine doublet with $F$ = 1/2, 3/2.  SI on the 6$^1S_0$ $\rightarrow$ 6$^3P_1$ transition leads to a small fractional change in the $E1$ absorptivity $\alpha$ for both the $F$ = 1/2 and 3/2 hyperfine components \cite{Lamoreaux1992}:
 \begin{equation}\label{Seq1}
 (\delta\alpha / \alpha)_{1/2} = - 2(\delta\alpha / \alpha)_{3/2} = a_{SI} (\hat \epsilon \cdot \vec{E})(\hat k \times \hat \epsilon) \cdot \vec{\sigma}
 \end{equation}
 where the SI amplitude $a_{SI}$ = $a_{M1} + a_{E2}$ is the sum of induced $M1$ and $E2$ contributions, and $\hat{\epsilon}$ and $\hat{k}$ are unit vectors for the light polarization and propagation direction, respectively. $\vec\sigma = 2<\vec I>$ is the ground state nuclear spin polarization. In the present study, $\sigma$ lies along an external magnetic field.

Both Eq. (\ref{Seq1}) and a related expression for SI-induced ground-state energy shifts are discussed below. We have used these shifts, manifested as $\vec{E}$-correlated modulations of the ground-state Larmor frequency $\Omega_L$, to measure $a_{SI}$. The measurements span several vector arrangements with $(\hat{\epsilon} \cdot \vec{E})(\hat{k} \times \hat{\epsilon})\cdot\vec{\sigma}$ $\neq$ 0 and together give:
\begin{equation}\label{SInum}
a_{SI} = (5.8 \pm 1.5)\times10^{-9} \hspace{2mm} \mbox{(kV/cm)}^{-1} \nonumber
\end{equation}
Separate measurements with $(\hat{\epsilon} \cdot \vec{E})(\hat{k} \times \hat{\epsilon})\cdot\vec{\sigma}$ = 0 give:
\begin{equation}
(\delta\alpha / \alpha)_{Null} = (0.6 \pm 1.8)\times10^{-9} \hspace{2mm} \mbox{(kV/cm)}^{-1} \nonumber
\end{equation}
For both, the quoted error is a quadrature sum of statistical and systematic errors; as shown below, the statistical errors dominate. The measured $a_{SI}$ agrees with the Ref. \cite{Beloy2009} prediction, $a_{SI}$ = 8.0$\times$10$^{-9}$ (kV/cm)$^{-1}$ within 1.5-$\sigma$. Moreover, $(\delta\alpha / \alpha)_{Null}$ is consistent with and provides a useful check on the expected Eq. (\ref{Seq1}) vector dependence.

Eq. (\ref{Seq1}) is obtained by evaluating the product of the $E1$ amplitude and the Stark induced $M1$ and $E2$ amplitudes.  The vector $\hat{\epsilon}$ comes from $E1$, $\vec{E}$ from the Stark mixing matrix, and $\hat k \times \hat \epsilon$ from the optical magnetic field and electric field gradient that drive the $M1$ and $E2$ amplitudes respectively.  The exact vector expression in Eq. (\ref{Seq1}) is derived rigorously in Ref  \cite{Hodgdon91} .
The absorptivity change $\delta \alpha$ due to each hyperfine line must be equal in magnitude and opposite in sign, as a nuclear spin-dependent effect must vanish if the hyperfine structure is not resolved.  Since the absorptivity $\alpha$ of the $F=3/2$ line is twice that of the $F=1/2$ line, this requires $(\delta \alpha/\alpha)_{3/2}=-(1/2)({\delta \alpha}/{\alpha})_{1/2}$ as in Eq. (\ref{Seq1}).

SI produces an energy shift $\delta U$ in the ground state magnetic sublevels and a change $\delta\Gamma$ in the photon absorption rate $\Gamma$.  We can relate $\delta\Gamma$ to the absorptivity change via $\delta\Gamma/\Gamma=\delta \alpha/\alpha$.   $\Gamma$ and $\delta\Gamma$ have a Lorentz lineshape $\gamma^2/(4\Delta\omega^2+\gamma^2)$ with $\Delta\omega$ the detuning from resonance and $\gamma$ the resonance FWHM, while $\delta U$ has a dispersion shape and is related to $\delta\Gamma$ by the dispersion relation \cite{Corney2006}:
\begin{equation}
\delta U_F = \hbar \frac{\Delta\omega_F}{\gamma_F}\delta\Gamma_F = \hbar \frac{\Delta\omega_F}{\gamma_F} \Gamma_F (\delta\alpha/\alpha)_F,
\end{equation}
where the subscript $F=\{1/2,3/2\}$ denotes the contribution of each hyperfine component individually.
The work presented here uses a laser \cite{Harber2000} tuned midway between the two hyperfine components. In this case, the shift to $\Omega_L$ in an external magnetic field $\vec{B}$ is given by
\begin{equation}\label{eq:SI4}
\delta \Omega_L =
\sum_{F=1/2}^{F=3/2} \frac{2\, \delta U_F}{\hbar} =
\frac{\Delta}{\gamma}\left[\frac{2}{3}\Gamma_{tot}\left(\frac{\delta \alpha}{\alpha}\right)_{1/2}\right],
\end{equation}
where $\Delta$ is the hyperfine splitting and $\Gamma_{tot}=\Gamma_{1/2}+\Gamma_{3/2}$ is the experimentally relevant total photon absorption rate; we have used $\Gamma_{{3/2}}$ = 2$\Gamma_{{1/2}}$ and the fact that $\delta\Omega_L$ is twice the shift of an individual magnetic sublevel. The Eq. (\ref{Seq1}) spin dependence allows $\delta \Omega_L$ to be modeled as a Zeeman shift induced by a virtual magnetic field along $\hat{k} \times \hat{\epsilon}$. Hence, shifts linear in $\vec{E}$ arise if $\hat{\epsilon}$, $\vec{E}$, and $\vec{B}$ are aligned such that $\hat k \times \hat \epsilon$ has a finite projection along $\vec{B}$ and $\hat \epsilon$ has a finite projection along $\vec E$.\cblack

\begin{figure}[t]
\includegraphics[width=\columnwidth, trim=0.25in 0.5in 0.25in 0.1in]{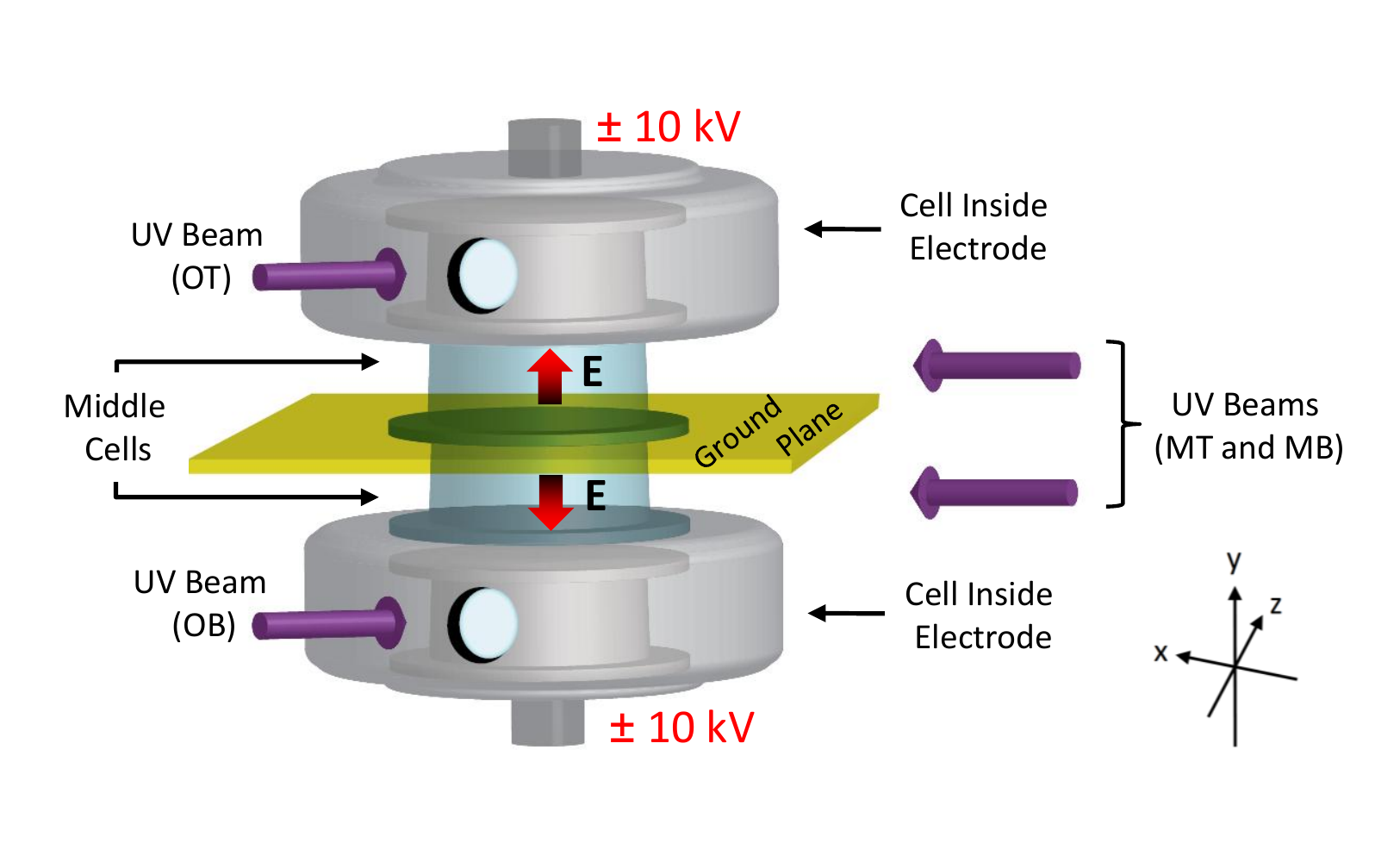}
\caption{(color online). Simplified diagram of the apparatus. OT, outer top; MT, middle top; MB, middle bottom; OB, outer bottom. The $\hat{k}$-vectors for the middle and outer beams point along $\hat{x}$ and $\hat{z}$, respectively. The $E$-field is along $\pm$ $\hat{y}$}\label{ExpD}
\end{figure}

Fig. \ref{ExpD} is a simplified diagram of the apparatus. Detailed descriptions appear elsewhere \cite{Griffith2009,Swallows2009}. Four isotopically enriched $^{199}$Hg vapor cells are placed in a uniform $B$-field. The middle cells have oppositely directed $E$-fields, giving SI-sensitive Larmor frequency shifts of opposite sign. The outer cells, placed at $E$ = 0 inside the high voltage (HV) electrodes, are free of SI effects and instead allow cancelation of $B$-field gradient noise and tests for spurious HV-correlated $B$-field shifts. The cell end caps are coated with SnO and act as $E$-field plates separated by 11~mm. The cells contain 475 torr of CO buffer gas and have paraffin wall coatings that enable typical transverse spin coherence times T$_2$ of 100--200 sec. The cells sit in an airtight, conductive polyethylene vessel housed inside three layers of magnetic shielding. To minimize leakage currents, the vessel is filled with 1 bar of N$_2$ which is flushed continuously.

\begin{figure}[rb]
\includegraphics[width=\columnwidth, trim=0.25in 0.in 0.25in 0.25in]{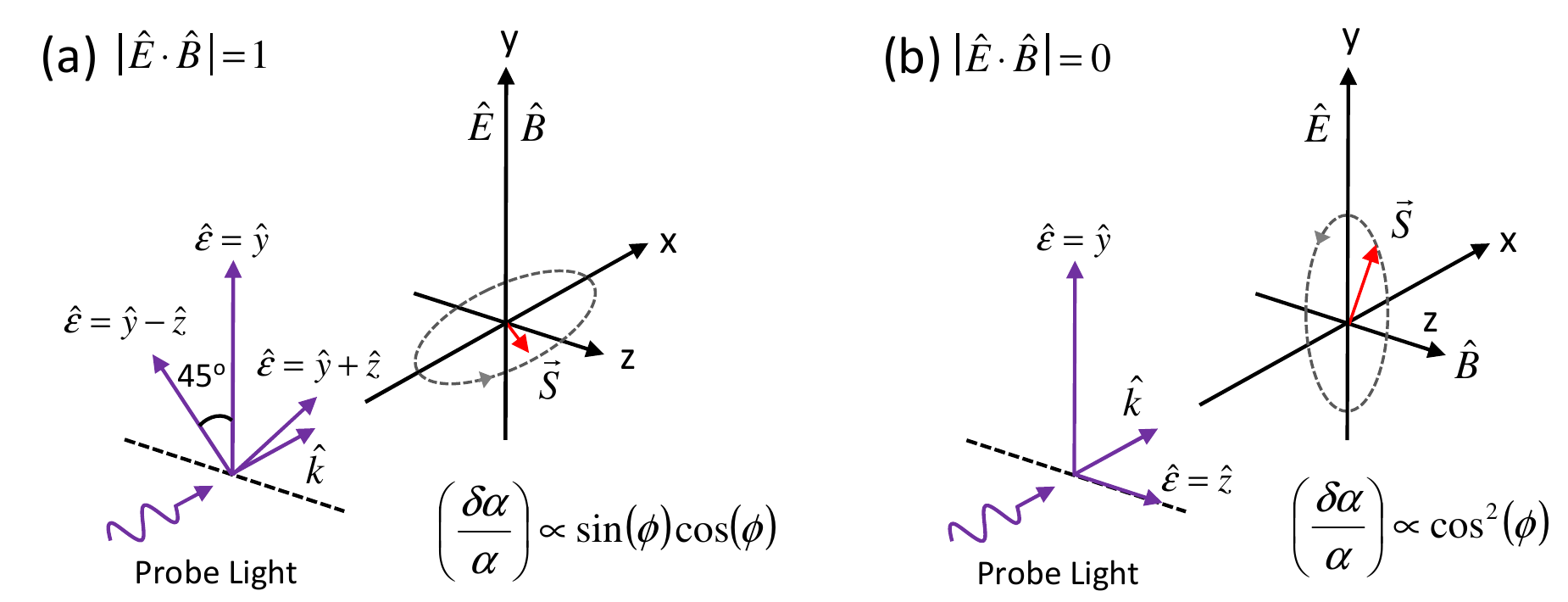}
\caption{Coordinate system for measuring $(a_{M1} + a_{E2})$ when (a) $\hat{E}\cdot\hat{B}$ = 1 and (b) $\hat{E}\cdot\hat{B}$ = 0.}\label{vectorD}
\end{figure}

The interference amplitude and $(\delta\alpha / \alpha)_{Null}$ were each measured for $|\hat{E}\cdot\hat{B}|$ = 1 and $|\hat{E}\cdot\hat{B}|$ = 0. The former (latter) used all four vapor cells (the middle cells: here, $\hat{k} \times \hat{B}$ = 0 in the outer cells, leading to inefficient pumping). Figure \ref{vectorD} shows the relevant coordinate systems. For Fig. \ref{vectorD}(a), the Eq. (\ref{eq:SI4}) angular dependence is $\sin{\phi}\cos{\phi}$ where $\phi$ is the angle between $\hat{\epsilon}$ and $\vec{E}$. In this case, $\hat{\epsilon}$ was set to either $\phi$ = 45$^\circ$ or -45$^\circ$, leading to SI-signals of $\delta\Omega_L$/2 or -$\delta\Omega_L$/2, respectively. Separately, $\hat{\epsilon}$ was set to $\phi$ = 0 where the expected shift is zero. For Fig. \ref{vectorD}(b), the angular dependence is $\cos^2{\phi}$. Here, measurements were taken with $\phi$ = 0 and $\phi$ = 90$^\circ$, leading to expected shifts of $\delta\Omega_L$ and zero, respectively. To test the Eq. (\ref{eq:SI4}) probe light intensity dependence, complete groups of data were taken for $\Gamma_p$ between $\sim$ 1/600 s$^{-1}$ and 1/50 s$^{-1}$.

Single experimental cycles (termed scans) use a pump-probe sequence; throughout, intensity stabilized 254 nm laser light enters each cell with $\hat{k}$ normal to the precession axis along $\vec{B}$. During the 30 sec pump phase, synchronous optical pumping with circularly polarized light tuned to the $^{199}$Hg $^1S_0(F=1/2) \rightarrow \, ^{3}P_1 (F=1/2)$ transition builds up spin polarization $\vec S$\cblack\ in a frame rotating about $\vec{B}$. During the probe phase, the light polarization is switched to linear and the frequency tuned midway between the $F=1/2$ and 3/2 hyperfine lines. The precessing $\vec S$ modulates the light polarization angle at $\Omega_L$. This angle is measured, for each cell, with a photodiode after a linear polarizer. The spin precession is monitored for 100--200 sec, after which the pump-probe cycle repeats. The HV is ramped to a new value during the pump phase, typically alternating between $\pm$ 10~kV.

The Larmor frequencies are determined by fitting the photodiode signals with exponentially decaying sine waves. Linear combinations of the four frequencies, $\nu_{OT}$, $\nu_{MT}$, $\nu_{MB}$, and $\nu_{OB}$ are then constructed. Here, OT is the outer top cell, MB is the middle bottom, etc. For $|\hat{E}\cdot\hat{B}|$ = 1, $\nu_{c} = (\nu_{MT} - \nu_{MB}) - \frac{1}{3} (\nu_{OT} - \nu_{OB})$ has the highest SI sensitivity since it maximally suppresses magnetic gradient noise (through 2nd order). For $|\hat{E}\cdot\hat{B}|$ = 0, the useful combination is $\nu_{m} = (\nu_{MT} - \nu_{MB})$. The SI signal, $\Delta\nu_{SI}$, is obtained from the HV-correlated component of $\nu_c$ or $\nu_m$ via 3-point string analysis \cite{Swallows2009}. Data runs last $\sim$ 24 hours and comprise several hundred scans. The run-averaged statistical error for $\Delta\nu_{SI}$ is set by the weighted error of the mean multiplied by the square root of the reduced $\chi^2$ where typically, $\chi^2$ $\leq$ 2. $\Gamma_p$ and $\phi$ were fixed during runs. $\phi$ was measured with calibrated polarizers, known to within $\pm$ 2$^\circ$, and changed daily between $\pm$45$^\circ$ (0, 90$^\circ$) for $|\hat{E}\cdot\hat{B}|$ = 1 ($|\hat{E}\cdot\hat{B}|$ = 0).

\begin{figure}[t]
\includegraphics[width=\columnwidth, trim=0.85in 1.4in 0.85in 1.2in]{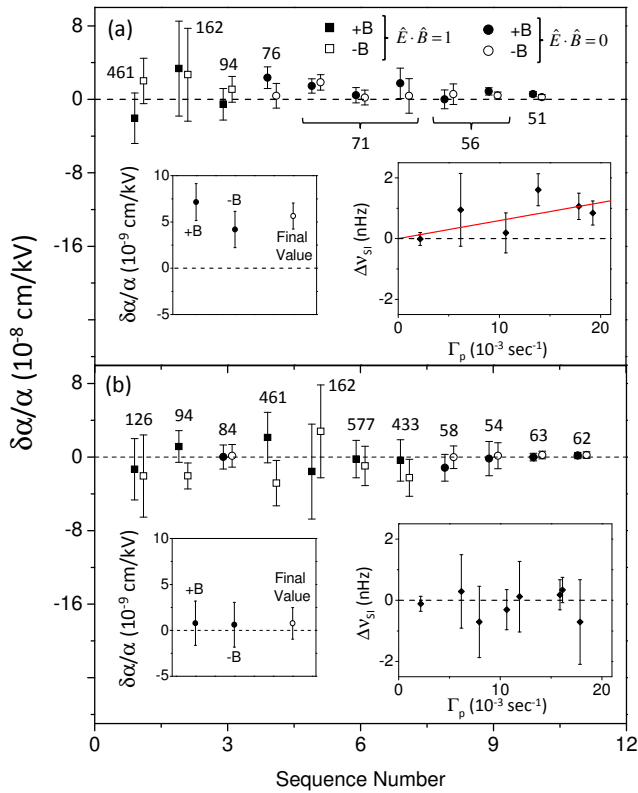}
\caption{Measured $\delta\alpha$/$\alpha$ for (a) $(\hat \epsilon \cdot \vec{E})(\hat k \times \hat \epsilon) \cdot \vec{\sigma}$ $\neq$ 0 and (b) $(\hat \epsilon \cdot \vec{E})(\hat k \times \hat \epsilon) \cdot \vec{\sigma}$ = 0. Numbers next to the points are sequence-average values for 1/$\Gamma_p$ in sec. The left-hand insets give the dataset-wide $+B$, $-B$, and final values for $\delta\alpha$/$\alpha$. The right-hand insets show the measured $\Delta\nu_{SI}$ versus $\Gamma_p$. The solid line in (a) is Eq. (\ref{eq:SI4}) with the measured final value for $a_{SI}$.}\label{dfig1}
\end{figure}

The ratio of residual circular to linear polarization for the probe beams, set mainly by cell wall birefringence and defects in the probe beam waveplates, was typically $<$ 0.13. At this level, derived values for $\delta\alpha$/$\alpha$ are negligibly impacted; e.g., a ratio of 0.3 produces a 2$\%$ error. The finite circular polarization, however, generates HV-independent vector light shifts ($\propto$ $\hat{k}\cdot\hat{B}$) whose fluctuations, due to scan-to-scan changes in $\hat{k}\cdot\hat{B}$ or $\hat{\epsilon}$, can lead to excess noise. For moderate probe intensities and the nominal 90$^\circ$ $\pm$ 0.5$^\circ$ angular alignment between $\hat{k}$ and $\hat{B}$, this noise was often negligible. Achieving similar performance at the highest intensities, however, required setting $\hat{k}$ to within $\delta\theta$ $<$ 0.1$^\circ$ of $\hat{k}\cdot\hat{B}$ = 0. To this end, for each cell, and prior to each high intensity run, $\hat{k}$ was set such that differential shifts due to flipping the probe polarization from right to left circular indicated $\delta\theta$ $<$ 0.1$^\circ$.

The probe intensity noise at $\Omega_L$ is typically 1.5$\times$ the shot-noise limit \cite{Swallows2009}. High intensity $|\hat{E}\cdot\hat{B}|$ = 0 runs used the normalized difference between the polarizer outputs for each cell. This step reduced run errors, on average, by 1.6$\times$, or roughly the expected factor of $1.5\sqrt{2} \sim 2$.

\cblack The measurement used four vapor cells, four electrodes, two vessels, multiple vapor cell and electrode orientations, and several permutations of the photodiode acquisition channels. We did not find statistically significant correlations between $\Delta\nu_{SI}$ and these changes. The components were altered between groups of 10-20 runs termed sequences; between sequences, the paraffin inside each cell was melted and the outer surfaces cleaned. Flips involving the vapor cells, electrodes, and vessels used nominally identical components. Each sequence included an equal number of SI-sensitive dipole HV runs (+ - + - HV sequence) for the two main $B$-field directions. Within a sequence, the HV ramp rate was permuted on adjacent runs between (4/n) kV/s where n = 1,2,4,6.

The measured SI amplitude was concealed by adding an unknown, SI-mimicking offset to $\nu_{MT}$ and $\nu_{MB}$ \cite{Griffith2009,Swallows2009}. This fixed blind was revealed only after the data collection, data cuts, and error analysis were complete.

The dataset comprised 181 runs: 47 used $|\hat{E}\cdot\hat{B}|$ = 1 while 134 used $|\hat{E}\cdot\hat{B}|$ = 0. In each case, a roughly equal number of runs involved non-null and null arrangements of $v$ = $(\hat \epsilon \cdot \vec{E})(\hat k \times \hat \epsilon) \cdot \vec{\sigma}$, with 99 (82) total runs taken for the former (latter). The statistical error for the entire dataset is 0.13 nHz, or within 2$\times$ of Refs. \cite{Griffith2009,Swallows2009}. Figure \ref{dfig1} shows the resulting sequence-level values for $\delta\alpha$/$\alpha$. For the expected null data in Fig. \ref{dfig1}(b), $v$ $\sim$ 0. Hence, Eq.~(\ref{eq:SI4}) gives $\delta\alpha$/$\alpha$ $\sim$ 0 independent of $\Delta\nu_{SI}$. To avoid this artificial zeroing, $\delta\alpha$/$\alpha$ in Fig. \ref{dfig1}(b) was calculated with $v$ set to the maximum allowed by $|\hat{E}\cdot\hat{B}|$; hence, the Fig. \ref{dfig1}(b) central values are upper limits.

\begin{figure}[b]
\includegraphics[width=\columnwidth, trim=0.85in 3.6in 1.5in 1.2in]{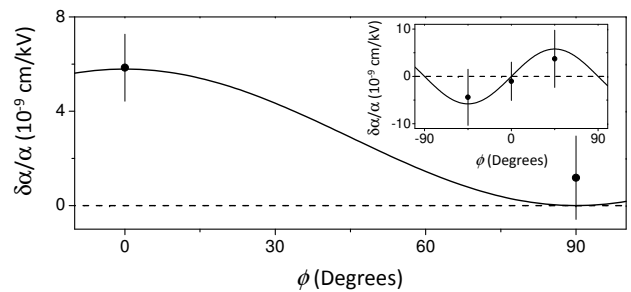}
\caption{$\delta\alpha$/$\alpha$ versus $\phi$ when $\hat{E}\cdot\hat{B}$ = 0. The inset shows similar measurements for $\hat{E}\cdot\hat{B}$ = 1. Solid lines give $\delta\alpha$/$\alpha$ versus $\phi$ predicted by Eq. (\ref{Seq1}) with the measured $(a_{M1} + a_{E2})$.}\label{dfig2}
\end{figure}

The sequence values are divided into one point for each $B$-field direction; each point is the weighted average of the relevant runs within the sequence. Numbers next to the points are sequence-average values for 1/$\Gamma_p$. In all cases, the $+B$ and $-B$ data are in good agreement. The weighted average of all the $+B$ and $-B$ data for the non-null and separately, null vector arrangements also agree within 1-$\sigma$. For both Fig. \ref{dfig1}(a) and \ref{dfig1}(b), $\delta\alpha$/$\alpha$ is also constant (within errors) over the $\sim$ 10$\times$ change in $\Gamma_p$ and thus consistent with Eq.~(\ref{Seq1}). In contrast, the right-hand insets show the expected $\Delta\nu_{SI}$ dependence on $\Gamma_p$: the Fig. \ref{dfig1}(a) slope from a least-squares linear fit constrained (not constrained) to pass through the origin is 63 $\pm$ 20 nHz/s (65 $\pm$ 14 nHz/s), while Eq. (\ref{eq:SI4}) and the measured $a_{SI}$ give 59 $\pm$ 15 nHz/s. For Fig. \ref{dfig1}(b), the slope for a linear fit constrained (not constrained) to pass through the origin is unresolved and equal to 8.5 $\pm$ 12 nHz/s (16 $\pm$ 26 nHz/s). For both Fig. \ref{dfig1}(a) and \ref{dfig1}(b), the fit intercept is unresolved at the 0.5-$\sigma$ level. Note that when $B$ is flipped, systematics that change sign relative to the SI signal can appear in $+B$, $-B$ differences, but will cancel in $+B$, $-B$ averages. Although the data are apparently free of such problems, sequence-level values were determined from straight $+B$, $-B$ averages.

Figure \ref{dfig2} shows $\delta\alpha$/$\alpha$ versus $\phi$ for $|\hat{E}\cdot\hat{B}|$ = 0. The central values for $\phi$ = 0 and $\phi$ = 90$^\circ$ are separated by $>$~2-$\sigma$ and agree with Eq. (\ref{Seq1}). The inset gives similar measurements for $|\hat{E}\cdot\hat{B}|$ = 1. Although differences between $\phi$ = 45$^\circ$, -45$^\circ$ and 0 are not resolved, $\delta\alpha$/$\alpha$ flips sign and passes through zero as predicted by Eq. (\ref{Seq1}).

Using the weighted mean of the sequence values: $a_{SI}$ = (5.8 $\pm$ 1.4$_{stat}$)$\times 10^{-9}$ (kV/cm)$^{-1}$ and $(\delta\alpha / \alpha)_{Null}$ = (0.6 $\pm$ 1.7$_{stat}$)$\times 10^{-9}$ (kV/cm)$^{-1}$. These final values include a numerically modeled, $\sim$ 3$\%$ correction due to the optical rotation-induced, time-varying projection of $\hat{\epsilon}$ onto $\hat{E}$. Note if $|\hat{E}\cdot\hat{B}|$ = 1 and $|\hat{E}\cdot\hat{B}|$ = 0 are considered separately, the relevant central values agree within 0.6-$\sigma$.

\begin{table}[t]
\caption[]{\label{Table1}
Systematic error budget (10$^{-10}$ cm/kV).}
\begin{tabular}{l@{\hspace{0.8cm}}c@{\hspace{1.4cm}}c@{\hspace{0.2cm}}} \hline\hline
Source & $\hat{E}\cdot\hat{B}$ = 0 & $\hat{E}\cdot\hat{B}$ = 1 \\\hline
Leakage Currents & 3.29 & 9.68 \\
Parameter Correlations & 2.91 & 16.6 \\
Charging Currents & 0.92 & 0.85 \\
E$^2$ Effects & 0.62 & 1.32 \\
Vector Alignment & 0.13 & 0.13 \\ \hline
\textbf{Quadrature Sum} & 4.53 & 19.3 \\
\hline\hline
\end{tabular}
\end{table}

Table \ref{Table1} summarizes the systematic errors. The {\it leakage current} error was conservatively estimated from worst-case scenarios for the single-cell currents: helixes for $|\hat{E}\cdot\hat{B}|$ = 1 and lines normal to $\hat{B}$ for $|\hat{E}\cdot\hat{B}|$ = 0. For the former (latter), the average single-cell current was 0.42 pA (0.5 pA). In both geometries, the $\Delta\nu_{SI}$ versus leakage current correlation slope was statistically unresolved. For $|\hat{E}\cdot\hat{B}|$ = 1, we use a 1/2 turn loop set by the cell geometry \cite{Griffith2009}, an effective current of $\sqrt{2}\times$0.42 pA = 0.59 pA (since fields in adjacent cells can add or subtract), and divide by 2 to account for averaging over the uncorrelated current paths in the four cells. A similar calculation was used for $|\hat{E}\cdot\hat{B}|$ = 0, but with the helical loop replaced by a $\sqrt{2}\times$0.5 pA = 0.71 pA line current normal to $\hat{B}$.

The {\it parameter correlation} error is the quadrature sum of $\delta p_i$ where each $\delta p_i$ is the product of the HV correlation for a given experimental parameter and the correlation of that same parameter with $\Delta\nu_{SI}$. Specific parameters are: the vapor cell spin amplitudes, lifetimes, relative phases, and UV transmission; the laser power, frequency, drive current, and control voltages; the three axis ambient magnetic field; and the $B$-field coil currents (main coil and up to three gradient coils). No statistically significant correlations were found. Under these conditions, the parameter correlation error is set largely by uncertainties in the correlations (and thus affected by the number of runs), leading to the larger value for $|\hat{E}\cdot\hat{B}|$ = 1.

The {\it vector alignment} error accounts for angular misalignment between $\hat{\epsilon}$, $\hat{k}$, $\hat{E}$, and $\hat{B}$ and scales as the product of these errors and the measured $a_{SI}$. The $E^2$ error for $|\hat{E}\cdot\hat{B}|$ = 0 is a 1-$\sigma$ upper limit for $(\delta\alpha / \alpha)_{Null}$ multiplied by the measured $<$ 2$\%$ $E$-flip asymmetry \cite{Swallows2009}. The remaining Table \ref{Table1} entries are given by Eq. (\ref{eq:SI4}) via errors in $\Delta\nu_{SI}$; the conversion uses an inflated misalignment angle of $\phi$ = $\pm$ 45$^\circ$ and dataset-wide means for $\Gamma_p$: 1/130 sec for $|\hat{E}\cdot\hat{B}|$ = 1 and 1/60 sec for $|\hat{E}\cdot\hat{B}|$ = 0. The $\Delta\nu_{SI}$ error analysis techniques are detailed in Ref. \cite{Swallows2009}.

Combining the relevant statistical and systematic errors in quadrature then gives $a_{SI}$ and $(\delta\alpha / \alpha)_{Null}$ for $|\hat{E}\cdot\hat{B}|$ = 1 and $|\hat{E}\cdot\hat{B}|$ = 0. Taking the error-weighted mean of these intermediate values:
\begin{eqnarray}
a_{SI} = (5.8 \pm 1.5)\times10^{-9} \hspace{2mm} \mbox{(kV/cm)}^{-1} \nonumber \\
(\delta\alpha / \alpha)_{Null} = (0.6 \pm 1.8)\times10^{-9} \hspace{2mm} \mbox{(kV/cm)}^{-1} \nonumber
\end{eqnarray}


This work was supported by NSF Grant PHY-0906494 and the DOE Office of Nuclear Science.


\newpage

\end{document}